\documentclass[sigconf]{acmart}
\usepackage{amsmath}

\usepackage{algorithm}
\usepackage{algorithmic}
\usepackage{multirow}
\usepackage{subfigure}
\usepackage{graphicx}
\usepackage{xcolor}
\usepackage{bm}
\newcommand{\tabincell}[2]{\begin{tabular}{@{}#1@{}}#2\end{tabular}}
\newcommand{\para}[1]{{\vspace{2pt} \bf \noindent #1 \hspace{1pt}}}
\AtBeginDocument{%
  \providecommand\BibTeX{{%
    \normalfont B\kern-0.5em{\scshape i\kern-0.25em b}\kern-0.8em\TeX}}}

\setcopyright{acmcopyright}
\copyrightyear{2018}
\acmYear{2018}
\acmDOI{10.1145/1122445.1122456}

\acmConference[Woodstock '18]{Woodstock '18: ACM Symposium on Neural
  Gaze Detection}{June 03--05, 2018}{Woodstock, NY}
\acmBooktitle{Woodstock '18: ACM Symposium on Neural Gaze Detection,
  June 03--05, 2018, Woodstock, NY}
\acmPrice{15.00}
\acmISBN{978-1-4503-XXXX-X/18/06}

\begin{document}

\title{MIRA:Leveraging Multi-Intention Co-click Information in Web-scale Document Retrieval using Deep Neural Networks}

%with graph attention network and Bert
\author{Yusi Zhang}
\email{Yusi.Zhang@microsoft.com}
\affiliation{
  \institution{Microsoft}
}
\author{Chuanjie Liu}
\email{chuanli@microsoft.com}
\affiliation{
  \institution{Microsoft}
}
\author{Angen Luo}
\email{Angen.Luo@microsoft.com}
\affiliation{
  \institution{Microsoft}
}
\author{Hui Xue}
\email{xuehui@microsoft.com}
\affiliation{
  \institution{Microsoft}
}
\author{Xuan Shan}
\email{xuanshan@microsoft.com}
\affiliation{
  \institution{Microsoft}
}
\author{Yuxiang Luo}
\email{Yuxiang.Luo@microsoft.com}
\affiliation{
  \institution{Microsoft}
}
\author{Yiqian Xia}
\email{yiqian.xia@microsoft.com}
\affiliation{
  \institution{Microsoft}
}
\author{Yuanchi Yan}
\email{Yuanchi.Yan@microsoft.com}
\affiliation{
  \institution{Microsoft}
}
\author{Haidong Wang}
\email{haidwa@microsoft.com}
\affiliation{
  \institution{Microsoft}
}
%
%\author{Lars Th{\o}rv{\"a}ld}
%\affiliation{%
%  \institution{The Th{\o}rv{\"a}ld Group}
%  \streetaddress{1 Th{\o}rv{\"a}ld Circle}
%\email{larst@affiliation.org}


\begin{abstract}
 We study the problem of deep recall model in industrial web search, which is, given a user query, retrieve hundreds of most relevance documents from billions of candidates. The common framework is to train two encoding models  based on neural embedding which learn the distributed representations of queries and documents separately and match them in the latent semantic space. However, all the exiting encoding models only leverage the information of the document itself, which is often not sufficient in practice when  matching with query terms, especially for the hard tail queries. Meanwhile, It has been proved that the click-through logs over query-document pairs in real search engine provide rich
information for multiple tasks in information
retrieval. Inspired by this, we aim to leverage the additional information for each document from its co-click neighbours to help document retrieval. The challenges include how to effectively extract  information and eliminate noise when involving co-click  information in deep model while meet the demands of billion-scale data size for real time online inference.

To handle the  noise in co-click relations, we firstly propose a web-scale Multi-Intention Co-click document Graph(MICG) which builds the co-click connections between documents on click intention level but not on document level, and it is scalable to billions of document nodes based real search engine logs. Then we present an encoding framework based on Bert and Graph Attention Networks(GAT)  which leverages a two-factor attention mechanism to aggregate neighbours and  can effectively handle the large amount of noise in the co-click relations. To meet the online latency requirements, we only involve neighbour information in document side whose vectors could be pre-built offline, and keep the query encoding only depends on its own text, which can save  the time-consuming  query neighbor search in real time serving. We conduct extensive offline experiments on both public dataset and private web-scale dataset from two major commercial search engines demonstrating the effectiveness and scalability of the proposed method compared with several baselines. And a further case study reveals that co-click relations mainly help improve web search quality from two aspects: key concept enhancing and query term complementary.

%categorizing and grouping the clicks of each document into different intentions when building the graph, and the two-factor attention mechanisms in the graph neural network, our proposed method can effectively handle the large amount of noise in the co-click graph. By restricting the co-click information only to the document side, we can avoid the expensive graph neighbor search time for online serving. We conduct offline experiments on both public dataset and private web-scale dataset from a major commercial search engine demonstrate the effectiveness and scalability of the proposed method compared with several baselines. And a online A/B test in production environment further shows our proposed model improve search engine revenue.
\end{abstract}


\keywords{web search, co-click graph, neural networks}


\maketitle

\section{introduction}\label{sec:intro}
In traditional web search, given a query, the documents are retrieved from large candidate corpus based on the relevance score which are usually some term based metrics such as BM25~\cite{robertson1995okapi}. Recently, with the fast rapid progress in neural text embedding and deep contextualized word representations such as ELMo ~\cite{peters2018deep} and BERT~\cite{devlin2018bert} , researchers begin to focus on applying deep neural network models to generate document and query embedding in latent space separately based on raw features  and   retrieve documents  according to their vector similarities. These methods improve the semantic understanding besides term match and can achieve better retrieve quality.

However,  in industrial  environment, due to the complexity of natural language and the diversity of user query expressions, web search retrieval  needs more input information for better document/query understanding and term matching, not just the promotion of the deep model architecture, especially for the long tail queries which only appear a few times or never before.
 However, almost all the existing deep models in IR only take the information of the document itself such as raw text as input.
They may provide a better query-document semantic understanding due to progress in model, but can't  provide additional information for matching.
Besides, ~\citet{guo2016deep} claim that Information Retrieval(IR) tasks are different from Nature Language Process(NLP) tasks, and that it is more important to focus on exact matching for the former and on learning text embeddings for the latter.
These inspires us to involve more information besides the query and documents themselves for better match and semantic understanding modeling.
Typically in industrial search engine, document information could come from several sources such as anchor, url, title and click,etc. Among them the click stream has been proved to be one of the most important features as it directly indicates user feedbacks. Thus it  provides us the feasibility to get potential  information for a specific document understanding from other similar documents through co-click relationship, e.g. two documents clicked by the  same query in history  indicates their similarity in some respects.

Using co-click information in  deep search model is not trivial.
Because it has too much noise, which is inconsistent with the need for  accurate information for term and semantic matching.
The noise in co-click relations mainly lies  in the following aspects:
%\begin{itemize}
%\item[*]False click. User false clicks may connect irrelevant queries and documents.
%\item[*]Multi-intentions. Users may click the same document with different search intentions. one document whose text is \textit{"amazing mens gothic t-shirts  black rose"}  may be clicked by users interested in two different intentions: \textit{"Gothic T-shirt"} or \textit{"Black rose T-shirt"}. As a result, the co-click documents involved from the click \textit{"gothic t-shirts"} would be noise for the query  \textit{"black rose t-shirts"}. This kind of multi-intention noise could be ignored by the traditional term match methods by nature, but can hurt the semantic deep encoding model if is not proper handled.
%\item[*]Semantic Difference. Even if two single-intention documents  have the same click, there might be differences in their semantics. E.g. two documents with themes "How high is Mount Everest" and "How many countries have climbed Everest" may have the same click "Mount Everest". Thus when encoding one document, completely introducing the text information of the other one will cause its vector erroneously shifting in the semantic space.
%\end{itemize}
\para{i)False clicks:}user false clicks may connect irrelevant queries and documents.
\para{ii)Multi-intentions:}one document may contain many aspects and users may only interest in specific part of them. As a result users may click the same document with different search intentions. E.g. one document whose text is \textit{"amazing mens gothic t-shirts  black rose"}  may be clicked by users interested in two different intentions: \textit{"Gothic T-shirt"} or \textit{"Black rose T-shirt"}.
Thus, the information introduced from a historical click may become noise  when the document is searched by a query belonging to another intention.
 As a result, the co-click documents involved from the click \textit{"gothic t-shirts"} would be noise for the query  \textit{"black rose t-shirts"}. This kind of multi-intention noise could be ignored by the traditional term match methods by nature, but can hurt the semantic deep encoding model if is not proper handled.
\para{iii)Semantic Difference:}even if two single-intention documents   have the same click, there might be differences in their semantics. E.g. two documents with themes "How high is Mount Everest" and "How many countries have climbed Everest" may have the same click "Mount Everest". But when encoding one document, completely introducing the text information of the other one will cause its vector erroneously shifting in the semantic space, and it need more feature extraction and noise eliminating.

In this paper, we focus on the web-scale web search problem in industrial environment and aim at enriching  document text through co-click relations to improve the retrieve quality. We first build the web-scale co-click graph based on real click log and extract neighbours for each document as complementary information. Then we  propose a siamese deep  model MIRA based on Bert and graph neural networks with a two-factor mechanism to encode queries and documents with their neighbours into continuous vectors. Our key challenge lies in two aspects:
i)how to effectively extract information from co-click graph with billions of nodes while eliminating the noises.
ii) how to scale both training and online serving of graph neural based embedding.

To handle the co-click noise problem, we firstly split the clicks of each document into different intention groups based on the term Jaccard Similarity and build the Multi-Intention Co-Click Graph, where each document is represented with several nodes and each node only contains clicks belonging to one intention. As a result, one node could only be reached through the same intention clicks thus avoiding the cross-intention information. Secondly, we propose to use graph attention networks with  a two-factor attention mechanism to help precisely extract neighbour information while eliminating false click and semantic difference noise. Specifically, we design a interaction attention factor for measuring vectors semantic correlation and another dot product factor measuring vectors term match correlation. As for the scalability challenge, instead of using query-document  bipartite graph, we only involve document nodes into our graph modeling, which alleviates the need for expensive real-time query neighbours search online. Furthermore, instead of using transductive training which need multiplying feature matrices by
powers of the full graph Laplacian, we use an inductive training method for GCN, which sample neighbour nodes for each document as neighbour subgraph, and thus dramatically reducing the training and inference cost.
%split the clicks of each document into different groups based on the token jaccard similarity. Then instead using one node to represent a document in co-click graph, we uses multi nodes
\begin{figure*}
\includegraphics[width=6in,,height=2.4in]{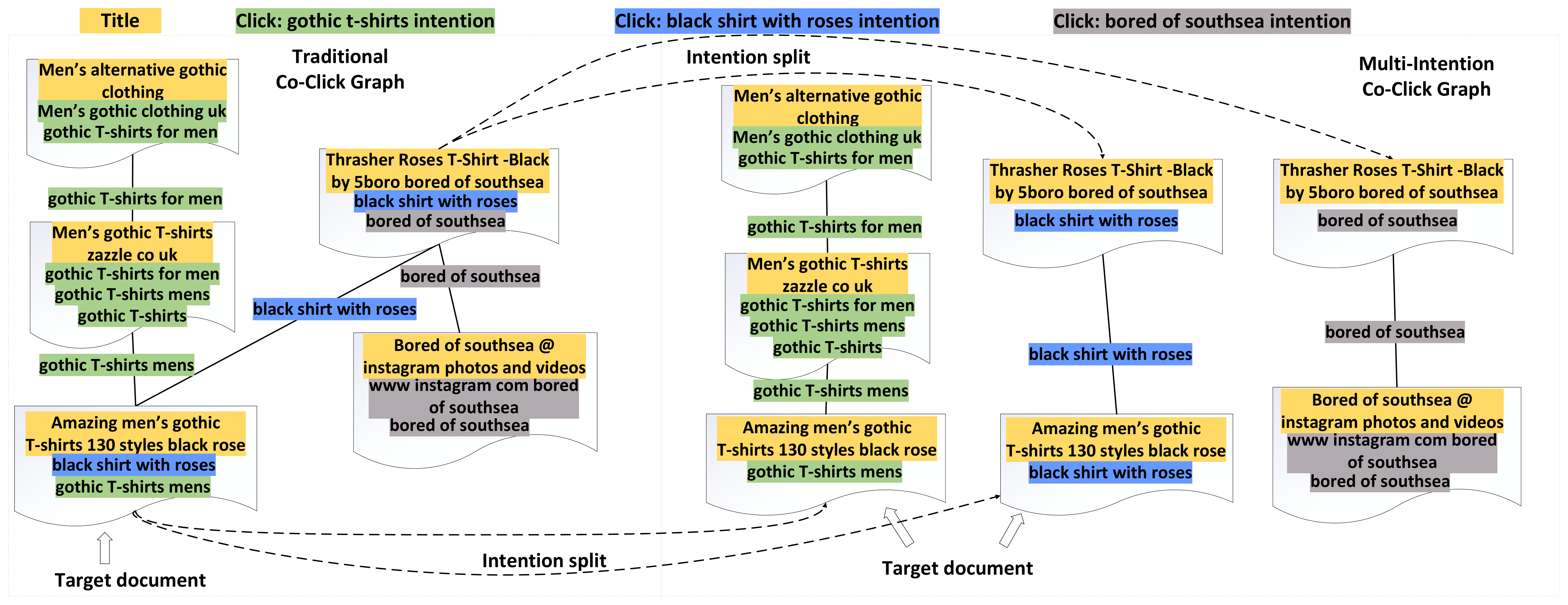}
\caption{The example of (left) traditional co-click graph and (right) multi-intention co-click graph for five documents, where we use different color to represent different source texts: yellow for title; green,blue  and grey for three different click intentions. %the solid line represents the co-click connection and the dotted line represent the document split.
From the figure we can see, no matter which intention of the target document the user is searching for, in traditional co-click graph we would introduce many irrelevant click intentions and  documents  when collect neighbours. However, in multi-intention co-click graph, we split document into fine-grained click intention level, and thus can  collect two set of neighbours that perfectly match target document's two click intentions respectively. }
\label{fig:CoClickGraph}
\label{fig:model}
\end{figure*}
We demonstrate the effectiveness and efficiency of our proposed framework on one public web search dataset collocted from SouGou and one private web-scale click data set collected from real search engine Bing. The offline experiments show our proposed model significantly outperforms various baselines.

Our contributions in this paper include:
\begin{itemize}
\item We take the first step to introduce additional document information for deep recall model in web search based on co-click relationships.
\item To handle the  noise in traditional click-bipartite graph, we proposed a novel multi-intention  co-click graph based on click intention level to get more accurate information connections.

\item We propose MIRA, a siamese deep model based on Bert and graph attention networks, which makes full use of the information of the document and its co-click neighbors while meeting the requirements of industrial online service. Besides, we design a two-factor attention mechanism for neighbour aggregation by measuring both semantic and term match correlations to address the false click and semantic difference problem.

\item We conduct extensive experiments on two real web-scale dataset(one from SouGou and one collected from Bing), and the results demonstrate the effectiveness of our proposed compared with several baselines. We then conduct a case study revealing that the co-click relations mainly help improve web search quality from two aspects: key concept enhancing and query term complementary.
\end{itemize}

\section{Related-work}
Recently more and more deep neural network models have been applied to web search targeting for better semantic matching~\cite{guo2016deep}~\cite{hu2014convolutional} beyond the term-based method such as BM25 ~\cite{robertson2009probabilistic}. ~\citet{guo2016deep} use  two separate feed forward network encoders to get query and document vectors based on their raw text. ~\citet{huang2013learning} adopt the similar model while they also introduce word hash method to reduce the bad-of-words vocabulary size. Researchers also studied different neural network models, e.g. ~\citet{salakhutdinov2009semantic} extend the LSA model by using a deep network (autoencoder) to discover the hierarchical semantic structures embedded in the query and the document.~\citet{hu2014convolutional} propose to use Convolutional neural networks(CNN) as the encoder to better capture  the rich matching patterns at different levels.

With the help of deep contextualized word representations and self-attention, two classes of fine-tuned architecture are typically built for sentence-pair embedding: Siamese-encoders and Cross-encoders~\cite{humeau2019poly}. To be specific in web search, Cross-encoders performs full  self-attention over a given query-document pair and thus will be extremely time consuming when it comes to industry where each query need to search in billions of documents. On the contrary, Siamese-encoders perform self-attention
over the query and document separately and calculates the cosine similarity with each final representation. It provides the feasibility for real time online search as we can reuse the pre-builded  document embedding. ~\citet{reimers2019sentence} firstly propose a siamese  Bert-based model and get the state-of-the-art  semantic textual similarity on some sentence pair embedding benchmarks.
However, all these works only take the text of the document itself as input
%As a result, it requires high semantic modeling ability of the encoder
and often suffer from the different expressions in document and query when model fail to match the in semantic space.

Mining query and document similarities from a  click-bipartite  graph has also been proposed by researchers.
~\citet{jiang2016learning} build the click-document bipartite graph and propose a propagation approach to learn a vector for each document. Click-document bipartite graph is also used by ~\citet{wu2013learning} where they use the matrix factorization method to generate query and document embedding from this graph adjacency matrix. ~\citet{craswell2007random} evaluate different random walk methods on click-document  graph and propose a Markov Random walk model to find relevant documents for each query in graph. ~\citet{beeferman2000agglomerative} propose to use this graph to identify query-query similarity in order to do query clustering. However, most of the existing works have some drawbacks. Firstly, they all lack the effective methods to handle all kind of noise in click-graph. Secondly, they all build the click graph as a  bipartite graph and thus their query modeling is depend on document neighbours which means it need to do real time neighbour search in a billion node graph when online serving, which is extremely difficult. Thirdly, some  methods are in a transductive manner. They focus on using the existing click-document pair and calculate the similarities among them, which can not handle the quickly updated documents and queries in industry. Finally, all of them use lightweight models like label propagation or random walk, which suffer the capacity loss compared to deep neural networks model. Besides handling all the drawbacks above, in this paper we don't directly get query-document similarity from the graph, instead we get more text information from the graph for better downstream encoding. ~\citet{xue2004optimizing} also use iterative algorithm to annotate association queries in click graph, but they only use them to help term match in inverted index while don't leverage their semantic information and  thus don't need pay attention to the noise problem. While we introduce more text including all the sources from other documents, as a result, we can involve more information and need more power to handle the noise, especially for semantic matching.

Graph Convolutional Networks(GCN)~\cite{kipf2016semi}, which the core idea is to aggregate neighbour features to learn topology structure as well as content information, has been proved to success in many area, such as recommendation~\cite{ying2018graph}, NLP~\cite{yao2019graph} and social network analysis~\cite{qiu2018deepinf}.
There is very few work to apply it in web search, ~\cite{scarselli2005graph} use graph neural network combining with the PageRank~\cite{brin1998anatomy} to rank web pages based on anchor relations. Their method relys on whole graph computation and is not scalable. To the best of our knowledge, we are the first to leverage GCN to incorporate co-click document information  to improve retrieve quality in industry large-scale web search scenario.

\section{Multi-Intention Co-Click Graph}
In this section, we first introduce our preliminaries and notations, then we  introduce the necessary and how to build the multi-intention Co-Click Graph.

\subsection{Preliminaries and Notations}
Let $q$ denotes the user query. And for each document $d$, we use the concatenation of its click stream $c$ and raw text  from other source $o$ to represent, e.g. $d=[o,c]$. Note here for each document , its click stream may contain several clicks where each click  represents one history query-and-click for this document. A document may contain many aspects of the content, and users searching and clicking  it may be because they are only interested in part of the content. As a result, clicks for one document may belong to different search intentions, referred as multi-intention issue. We use $c^{i,j}$  to detail represent document clicks where $i$ represents the $i-th$ click  and $j$ denotes this click  belongs to the $j-th$ intention of this document,$c=[c^{1,1},...,c^{i,j}]$.

%\begin{figure}
%\subfigure[Traditional Co-Click Graph]{\label{fig:TraditionalCoClickGraph}
%\includegraphics[width=1.6in, height =1.5in]{fig/TraditionalCoClickGraphNew.pdf}}
%\subfigure[Multi-Intention Co-Click Graph]{\label{fig:MultiIntentionCoClickGraph}
%\includegraphics[width=1.4in, height =1.5in]{fig/MultiIntentionCoClickGraphNew.pdf}}
%\caption{The example of (a) traditional co-click graph and (b) multi-intention co-click graph for five nodes. In multi-intention co-click graph the dash line represent the virtual nodes and links just for better understanding. We display all the 2-hop neighbour texts each graph aggregates for A and label the three different texts corresponding to document A's first intention, second intention and irrelevant to A with green red and yellow.}
%\label{fig:CoClickGraph}
%\end{figure}

\subsection{Traditional Co-Click Graph}
Based on the full document click stream , we can build co-click graph for all the documents. For traditional co-click graph $G(V,E)$, each node $v_i$ represents one document with the full text as its node features, each edge $e_{ij}$ represents $v_i$ and $v_j$ have at least one common click.

However,  each document may have some clicks that belong to different intentions, while the user query is often very specific. As a result the traditional co-click graph may involve much noise if  neighbour documents are taken as features due to its coarse-grained nodes. As shown in Figure ~\ref{fig:CoClickGraph}, we have five documents  and each of them contains one title and may have several clicks that may belong to different intentions. When encoding the target document, if we take all its 2-hop neighbours into consideration, we can get all the nodes appeared in left part of Figure ~\ref{fig:CoClickGraph} as complementary information. However, if a user searches \textit{"T-shirt with roses"}, although the target document is one perfect match result for this query, the neighbour documents which connected to it through click \textit{"gothic T-shirts mens"} will only provide irrelevant information w.r.t the query.  Moreover, this multi-intention noise will rapidly enlarge with the expansion of the neighbor's hop. E.g. under the above setting, the document titled \textit{"Thrasher Roses T-shirt -Black by 5boro bored of southsea"} may provide some useful information because of the high similarity to target document and query. But it's another neighbour, which is  reached through click \textit{"bored of southsea"}, will be irrelevant w.r.t the query.  All these irrelevant neighbour information will do harm to the semantic encoding.
\begin{figure*}
\includegraphics[width=7in,,height=3.in]{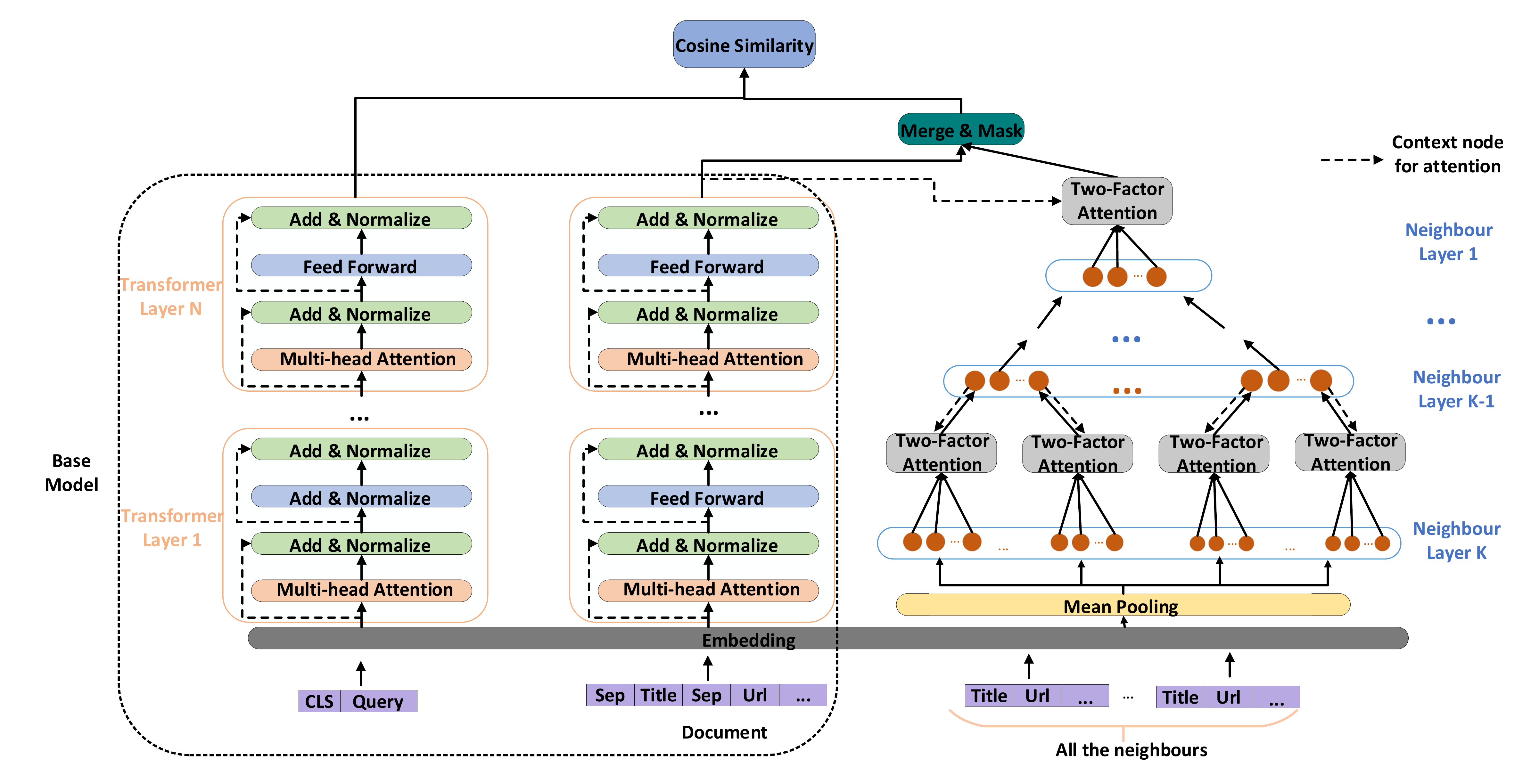}
\caption{The Model Architecture of MIRA.}
\label{fig:model}
\end{figure*}

\subsection{Multi-intention Co-Click Graph} To handle the multi-intention noises in traditional co-click graph, we propose the Multi-Intention Co-Click Graph(MICG) for a better neighbour connection. The key idea of the MICG is to characterize and divide the clicks of each document into different intention groups, then we build the graph based on click intention group level but not on document level.  In MICG each node only contains one intention group of click together with other raw text  of the corresponding document, and each edge $e_{ij}$ now represents that $v_i$ and $v_j$ have at least one common click item that belongs to their respective intention groups. E.g.  As shown in right part of Figure ~\ref{fig:CoClickGraph}, the target document is now split into 2 nodes and if we get its neighbours from these two nodes separately, we can get two set of neighbours whose text information perfectly matching  two click intentions, \textit{"gothic T-shirts mens"} and \textit{"T-shirt with roses"} respectively.

\para{Click group algorithm.} We adopt the token-based Jaccard Similarity ~\cite{levandowsky1971distance} to split clicks into different intention groups. To be specific, for each document, we firstly sort  all its clicks based on the click importance, e.g. history click frequency. Then we tokenlize each click  into  a click token list $c_{token}$ using WordPiece ~\cite{wu2016google} to address  the misspelling problem. Next, we adopt an iterative method to group click token lists. In each loop, we select the currently highest ranked ungrouped click token list $c_{token_i}$ as main list and calculate the jaccard similarity $\zeta_{ij}$ between the remaining ungrouped click token lists $c_{token_j}$ and it.
\begin{equation}
\zeta_{ij} = \frac{c_{token_i} \cap c_{token_j} }{c_{token_i} \cup c_{token_j}}
\end{equation}
 By the end of this loop, we combine all the clicks whose token jaccard similarity with main click is higher than the threhold $\xi$ together with the main click  as one click intention group. We iterate this process until all the clicks are grouped or the intention group number has reached the upper bound for this document.

Then to fully leverage the group clicks in MICG, in training and evaluation, we build one vector for each node based on  their node features and neighbour information. As a result, we can get several vectors for each multi-intentions document, which emphasize different search intentions.

\section{Mira Framework}

In this section, we formally propose our encoding model MIRA based on  Bert and graph convolution network with a two-factor attention mechanism, which follows the siamese architecture  to independently derive query and document embedding. To encoding one document in co-click graph, one straightforward way is to concatenate both its raw text and neighbours'  into one lone sequence and apply it to the existing deep model. However it has two drawbacks, firstly it can not extract neighbour topology features. Secondly, typically one document contains  hundreds of words, so the sequence length after concatenating with neighbors may be too long for deep models (Bert has a length limit of 512). Although some works~\cite{yang2019xlnet} are proposed to handle the super long sequences, they are too time-consuming for encoding billions of documents. The model we proposed uses the BERT to perform  self attention on the document's own text, and uses GCN to process the information of its neighbors with relative small overhead . It can fully extract the topology features in the  MICG, while achieving good balance of efficiency and effectiveness.
Next, we firstly introduce our  base model and then describe how we incorporate graph convolutional networks into the model, and finally present our training strategy.

\subsection{Base model}
We adopt a siamese architecture just like ~\cite{reimers2019sentence} due to the scalability consideration. However, their model only targets for sentence embedding while in web search each document may have text from different sources such as anchor,title,url and body that need to be distinguished. So we further extend the model for multi source document embedding which will be introduced later.
As shown in the left dotted box in  Figure ~\ref{fig:model}, The base model mainly consists of 3 components: a shared embedding layer, a Bert based query encoder and a Bert based document encoder  .

\para{Shared embedding layer.}
In order to fully leverage word semantic information, both query and document encoder share the same embedding layer. We adopt the  WordPiece embeddings. Query and document texts are firstly tokenlized with WordPiece. For a given token, its final embedding representation is constructed by summing the corresponding token,segment and position embeddings.

%Then to form the final input token sequence, we add a leading $[CLS]$ token before the text tokens and use $[SEP]$  token to split different text tokens. Next the token sequences are fed into embedding layer to get their token embeddings.

\para{Bert-Based encoder}
Then we use two BERT-like model on top of the embedding layer, each of them contains several Transformer layers~\cite{vaswani2017attention}. For query side,  we add a leading $[CLS]$ token before the text tokens as its input sequence and use the  output of the $[CLS]$ token at last layer as the final embedding.
\begin{equation}
v_q = \varphi^q(q)
\end{equation}
Where $v_q$ denotes query embedding and $\varphi^q$ represent the token embedding and Bert encoding process for query.

As for document embedding, to handle the multi source texts, firstly we concatenate all the text tokens as input. Then except for the same leading $[CLS]$ as query side, we also add $[SEP]$ tokens between document tokens from different sources. Besides, we also assign different segment id for tokens from different sources. In this way, the model can better perceive tokens from different sources through different segment embedding. Finally we also use the $[CLS]$ token embedding as output.
\begin{equation}
v_d=\varphi^d(d)
\end{equation}
The two Bert models are both initialized from BERT-Base but they don't share parameters during training, and thus there are no dependency during inference.

\para{Similarity and Loss}
After get the  $v_q$ and  $v_d$ , we use the  cosine similarity $s$ between them as the relevance score of each query-document pair.
\begin{equation}
s = \iota_2(v_q)*\iota_2(v_d)
\end{equation}
Where $\iota_2$ denotes the l2 normalization.
 And we use the sigmoid cross entropy loss as the objective function.
\begin{equation}
L=-y\log(\sigma(\psi*s-\tau))-(1-y)\log(1-\sigma(\psi*s-\tau))
\label{equ:loss}
\end{equation}
Where $\sigma$ indicates the sigmoid function and $y$ is the relevance label for this query-document pair, while $\psi$ and  $\tau$ are two hyper-parameters.

\subsection{Graph Attention Network model}

To incorporate the neighbour texts into main document embedding, we propose to use graph attention network to extract features from its neighbours. We firstly describe how we build local neighbour subgraph for each node then introduce our forward convolve algorithm and the two-factor attention mechanism.

\para{Neighbour Subgraph.}
To meet the demands of web-scale document encoding, we adopt an inductive training method.  We firstly build calculation subgraph for each node. To be specific, For node $v$ in multi-intention  co-click graph, we firstly  sample $n$ nodes from its one-hot  neighbours $N(v)$ %together with  $v$
as the first layer neighbours , denoted as $S^1(v)$. Then for all nodes $v_i \in S^1(v)$, we also sample  $n$ nodes for  each from their one-hot  neighbours %together with $S^1(v)$
to  form $S^2(v)$. We can get at most $K$ layer neighbors in this way and all the neighbours $\sum_{i=1}^k S^i(v) $  are taken as  local neighbour subgraph $G_{v}$ for $v$. Note here we only take the one-hot neighbour in each layer due to the precise information requirements in information retrieval.

\para{Layer-Wise Attention Convolve.}
We now introduce how our model generate embedding for one node based on its local neighbours with a layer-wise  $\textit{Attention Convolve}$ operation

First of all, for each node $i$, we concatenate all the texts from different sources it contains as input. Then the text sequence is tokenlized and fed into the embedding layer. For those documents without graph neighbours, we pad all zeros vector as their neighbour embedding.  Next, considering the efficiency for model training and inference for web-scale data, we don't use the Bert to extract neighbour features, and instead we adopt a relative light way by using the mean pooling of all the token embedding as node embedding $t_i$. Note here all the $[SEP]$ and $[CLS] $ tokens used previously in Bert are removed as we adopt mean pooling instead of self-attention afterwards, and involving the embedding of these separators may affect the information of the node itself. While we still keep the different segment embedding for better understanding of the different sources. Then we further extract features from the representation with a dense neural network and $Gelu$~\cite{hendrycks2016gaussian} activation.
\begin{equation}
r_i=Gelu(\bm{W_h}*t_i+\bm{b_h})
\label{equ:representation}
\end{equation}
Where $\bm{W_h} \in \mathbb{R}^{F\times F}$ and $\bm{b_h}$ are trainable parameters, and $r_i$ is the node representation for node $i$.

Next, the graph neural network needs an aggregation step to collect neighbor information. %and a combination
%step that merges this information with node features.
Graph Attention Network  is a
recent proposed technique that introduces the attention mechanism to do neighbour aggregation. It performs self-attention on each node's neighbours to compute how important they are to that node.
Formally, given a node i, an important coefficient $\beta_{ij}$ is computed through a shared attentional mechanism $\textit{attn}:\mathbb{R}^{F} \times \mathbb{R}^{F} \rightarrow \mathbb{R} $ between node $i$ and node $j$  if $e_{ij}$ exists. Here we use $z_i$ and $z_j$ denote their current embeddings.
\begin{equation}
\beta_{ij} = \textit{attn}(\bm{W}z_i,\bm{W}z_j)
\label{equ:ori_atten}
\end{equation}
After that, to make coefficients easily comparable across
different neighbour nodes,  a softmax function is applied to normalize them
\begin{equation}
\alpha_{ij}=softmax_j(\beta_{ij})=\frac{exp(\beta_{ij})}{\sum_{v_j\in N(i)}exp(\beta{ij})}
\end{equation}

Once obtained the normalized attention coefficients, we can use them to compute a linear combination of the
feature vectors, to serve as the aggregated neighbour  vector  $h_{i}$ after applying a nonlinearity.
\begin{equation}
h_{i}=\sigma (\sum_{j \in N_i} \alpha_{ij}\bm{W_t}h_j)
%h_n_i=\sigma(\sum_{j \in N_i} \alpha_{ij}\bm{W}h_j)
\end{equation}

We then concatenate the aggregated neighborhood vector $h_{i}$ with the node's own embedding $z_i$
 and transform the concatenated vector through another dense
neural network layer  to get the new presentation $h_i^{new}$ of $i$
\begin{equation}
h_i^{new}= Gelu(\bm{W_c}[h_i,z_i]+\bm{b_c})
\label{equ:concat}
\end{equation}
We define equation~\eqref{equ:representation} - ~\eqref{equ:concat} together as \textbf{$\textit{AttentionConvolve}(h_i,N(i))$} operator.

\para{Multi-layer Convolve}
For each document, we stack multi \textit{Attention Convolve} layers corresponding to its neighbour subgraph.
As the aggregation iterates from the bottom layer(layer $K$) to the  top layer(layer 1) of its subgraph, the information of the bottom neighbor nodes  gradually converges toward the top nodes and the embedding incrementally gains more and more information about the local graph.  And finally all the  neighbour information  is accumulated in top layer neighbour node vectors.
Particularly, as shown in Figure ~\ref{fig:model}, the inputs to the convolutions at layer k depend on the vector output from layer $k +1$ except for the bottom layer, where the input are equal to the  node features $r_*$. Note
that the model parameters are
shared across the nodes in the same layer but differ between layers

\para{Two-factor multi head attention.}
An important innovation in our framework is  that we design an attention which is conducive to improving document retrieval.
In the original GAT, the $\textit{attn}$ in equation ~\eqref{equ:ori_atten} is  a single-layer feedforward neural network parametrized by a weight vector $c \in \mathbb{R}^{2F} $ and applying the LeakyReLU nonlinearity.
\begin{equation}
\beta_{ij} = LeakyRelu(\bm{c^T}([\bm{W}z_i,\bm{W}z_j]))
\end{equation}
where $.^T$ represents transposition.

However, in web search scenario, we argue that only using $\beta_{ij}$ as an indicator to represent neighbour correlation is not enough.
Intuitively,
%we need to using this attention mechanism to further eliminate the noise in co-click graph. Thus
we want to assign more weight to the document neighbours which are more correlation to the main document and thus may provide more useful additional information for matching while lowing the irrelevant neighbours' weight because they are high likely to introduce noise.
Basically speaking, the relevance between two documents can be measured in two aspects: term matching and semantic similarity.
As for $\beta_{ij}$,
it  concatenates the main vector and the neighbour vector and then  evaluates the correlation through  full dimensional vector interaction in the same latent space. This full interaction could be more helpful in getting the semantic similarity, but is relatively weak in evaluating the term matching. However, as introduced in sec~\ref{sec:intro}, sometimes exact term matching is more important in IR for evaluating similarities. So besides $\beta_{ij}$, referred as \textit{interaction factor}, we further propose \textit{dot-product factor} $\xi_{ij}$ to measure neighbour importance from term match perspective.
\begin{equation}
\xi_{ij} = Tanh((\bm{W_d}z_i*\bm{W_d}z_j))
\end{equation}
Where $Tanh$ represents the hyperbolic tangent function. The intuition is that if two documents have many same terms in their texts, then their vectors will be highly similar in some dimensions, although they may be different in other dimensions. Such partial dimensional similarity can be amplified and measured by dot-product~\cite{tata2007estimating}.

Next we combine these two correlation indicators together as a \textbf{two-factor attention mechanism}.
Fully expanded out, it can be formulated as :
\begin{equation}
\label{equ:full_attention}
\alpha_{ij}=\frac{exp(LeakyRelu(\bm{c^T}(\bm{W}z_i||\bm{W}z_j))+Tanh((\bm{W_d}z_i*\bm{W_d}z_j)))}{\sum_{v_j\in N(i)}exp(LeakyRelu(\bm{c^T}(\bm{W}z_i||\bm{W}z_j))+Tanh((\bm{W_d}z_i*\bm{W_d}z_j))}
\end{equation}

In addition, we also apply
\textbf{multi-head graph attention} as suggested by GAT. The multi-head attention mechanism performs
$M$ independent single attention in parallel, and aggregates the output of $M$ single attention
together through an aggregation function:
\begin{equation}
\label{equ:merge}
h_i=Aggregate_{m=1}^M(\sigma (\sum_{j \in N_i} \alpha^{m}_{ij}\bm{W_m}z_j))
\end{equation}
Here $Aggregate$ denotes the concatenation operator except for an mean pooling for the last layer.

And define ~\eqref{equ:full_attention} ~\eqref{equ:merge} as \textbf{$\textit{Attention}(h_i,N(i))$}.

\subsection{Intergrating Bert and Graph Attention Networks}

Compared to the one layer dense neural network, Bert performs self attention on the input sequence and thus has  more powerful feature extraction ability. To fully leverage Bert output, so for each main document $i$, after getting its first layer neighbour embedding $h_i^{1}$,
, we directly use  Bert encoder's output $v_i^{b}$ as the main document $i$'s own representation instead of the  $r_i$ generated from the equation ~\eqref{equ:representation} for better performance and generate the final document neighbour embedding $v_i^g$ with all the first layer neighbours attention it.
\begin{equation}
v_i^g=Attention(v_i^b,N(i))
\end{equation}
Note here although we only use the first layer node embeddings, they have already contain all the neighbour information due to the multi layer convolves.

Then for documents which has neighbour nodes, we combine the $v_i^{g}$  and $v_i^{b}$  through a weighted sum operation to get the final document vector. For those documents without co-click neighbours, although we use all-zero vectors as their neighbour embedding vectors, their $v_i^{g}$  are not all zeros due to the bias term introduced in the graph calculation process. So for this situation we mask the $v_i^{g}$ to all-zero vector to avoid damage to the final result of the model.
\begin{equation}
v_i=v_i^{b}+\lambda *v_i^{g}*\gamma
\end{equation}
Where $\lambda$ is a hyper-parameter, and $\gamma$ indicates whether this main document has co-click neighbours, $1$ representing yes and $0$ for no .

\begin{algorithm}[t]
\caption{Encoding Algorithm}
\label{alg:A1}
\begin{algorithmic}[1]
\REQUIRE One query $q$, one document $d$, One subgraph $G_d$ with $K$ layers, One hyperparameter $\gamma$.
\ENSURE The cosine similarity $s$ of this query document pair;
\STATE $h_i^{K} \Leftarrow r_i, \forall i \in S_d^K$ \\
\COMMENT{/* Generating neighbour vector for $d$              */}
\FOR{$k=K-1$ to $1$}
\FOR{$i \in S_d^k$}
\STATE $h_i^k=\bm{Attention Convolve}(r_i,(\cup h_j^{k+1}, \forall j \in N(i)))$
\ENDFOR
\ENDFOR\\
\COMMENT{/* Generating Bert output vector for $d$ and $q$    */}
\STATE $v_d^{b}= \varphi^d(d)$ \\
\STATE $v_q= \varphi^q(q)$ \\
\COMMENT{/* Intergrating Bert vector and neighbour vector for $d$      */}
\STATE $v_d^{g}=\bm{Attention}(v_d^{b},S_d^1))$
\STATE $v_d = v_d^{b}+\lambda*v_d^{g}*\gamma $ \\
%\COMMENT{/* Calculate the cosine similarity */}
%\STATE $s=\iota_2(v_q)*\iota_2(v_d)$
\end{algorithmic}
\end{algorithm}

The whole encoding algorithm is summarized in Algorithm~\ref{alg:A1}. We train our model in a mini-batch mode and use the same loss function ~\eqref{equ:loss} as base model.

\section{Experiments}
To demonstrate the efficiency and the quality of the
document retrievals of our proposed model, we conduct a comprehensive suite of
experiments based on large-scale real world search engine dataset  including offline
experiments,ablation study as well as case studies.

\subsection{Evaluation Dataset}
Our experiments are conducted on two real search engine  dataset, one public dataset the TianGong-ST  collected from Sougou.com and one private industrial-scale  dataset collected from Bing.com.

\para{The TianGong-ST Dataset~\cite{chen2019tiangong}:} This dataset is refined from an 18-day search log by Sogou, the second largest search engine in China. The dataset consists of 147,155 refined Web search sessions, 40,596 unique queries, 297,597 Web pages. Each document consists of title,body,url and clicks as raw texts. This dataset also provides 4000 human judged query-document pairs labeled  as ground truth.

\para{The  Bing dataset:} Our industrial experiment data were extracted from Bing's real logs. It totally contains 42.3 billion documents, among them 4.2 billion documents have history clicks, and besides clicks we also  use   title,anchor and url  as their raw texts.  Due to the large scale, we build the  co-click graph on these document, with the constrain that each document has maximum 5 click intention groups. As a result, we finally get a co-click graph with 6.9 billion nodes and 348 billion edges. As far as we know, this is the largest co-click graph  applied in web search research.

\begin{table*}[t]\caption{NCG results on Microsoft dataset}
\centering
\setlength{\tabcolsep}{1mm}{
\begin{tabular}{|c|c|c|c|c|c|c|}
  \hline
  &\multicolumn{4}{|c|}{Bing Dataset}& \multicolumn{2}{|c|}{Tiangong-ST Dataset}\\ \hline
  &\multicolumn{2}{|c|}{Head} &\multicolumn{2}{|c|}{Tail}&\multicolumn{2}{|c|}{ } \\ \hline
   & NCG@20&NCG@80 &NCG@20 &NCG@80 & NCG@5 &NCG@20\\ \hline
  Bm25  & $-27.61\%$	&$-20.59\%$ & $-27.87\%$ & $-21.70\%$ & $25.33$ & $52.63$ \\\hline
  VPCG & $-33.5\%$ &$-26.66\%$&$-31.10\%$& $-30.88\%$&$14.84$ & 41.03 \\ \hline
  Siamese Base model & $0.00\%$	&$0.00\%$ & $0.00\%$ & $0.00\%$ & $37.41$ & $62.44$ \\\hline
  MIRA-traditional & $+0.94\%$	&$+1.49\%$ & $+2.73\%$ & $+1.84\%$ & $41.68$ & $78.69$  \\\hline
  MIRA-NoAttention& $+0.41\%$	&$+0.42\%$ & $+1.22\%$ & $+1.18\%$ & $53.31$ & $81.89$ \\\hline
  MIRA-GAT& $+3.02\%$	&$+2.31\%$ & $+3.19\%$ & $+3.28\%$ & $55.43$ & $83.19$ \\\hline
  MIRA &$+4.11\%$ &	$+3.24\%$	&	$+5.10\%$ & $+4.66\%$ & $57.50$ & $87.62$ \\\hline
\end{tabular}}
\label{tab:off_result}
\end{table*}
\subsection{Evaluation Metrics}
As we targeting at the first step  retrieval in web search, we only focus on whether the high quality documents are in the recall list, but not on their specific rankings. Thus we adopt Normalized Cumulative Gain (NCG@inf) as our evaluate metric. For each query in validation and test set, we have at least one human-judged non-bad document in the candidate set, and we denote the non-bad document set w.r.t query $q$ as $\S_q$ .  Each non-bad document $d_i$  is labeled with $rel_i \in [1,4]$, where the larger $rel_i$ means the better relevance. So the NCG@k for query set $Q_n$ is defined as:
\begin{equation}
NCG@k = \sum_{q=1}^{n}\frac{\sum{2^{rel_i}-1,\forall i \in \nu_q^k}}{\sum{2^{rel_j}-1,\forall j \in \S_q}}
\label{equ:NCG}
\end{equation}
where $\nu_q^k$ is the non-bad document set which  exist in the top k recall list for query $q$.
To protect the privacy, we only report the relative improvement of our
proposed model against the baseline.

\subsection{Baselines}
We compared our model with the  feature based methods, graph based
methods, siamese deep model and our own model variants.

\para{BM25 based retrieval model.} We take BM25~\cite{robertson1995okapi} as an example of feature based methods. BM25 and its variants
have been extensively described and evaluated in IR literature, and hence serve as a strong baseline. We train a score function based on BM25 and other query-independent document features together with query features.

\para{Graph based methods.} Recently researchers mainly proposed M-PLS~\cite{wu2013learning} and VPCG~\cite{jiang2016learning} to leverage click-graph in document retrieval. The former use a Multi-view Partial
Least Squares method to learn a similarity function aims at maximizing the similarities of the observed query-document pairs on
the enriched click-through bipartite graph, it need to calculate SVD of click matrix and thus not scalable. The VPCG uses a propagation method to propagate vector information in click-document bipartite graph and calculate the similarity based on the final vectors. In this paper we choose VPCG as our graph based baseline due to the scalability consideration.

\para{Siamese Bert-based Model.} We use deep base model introduced in section XX, which has been achieved the SOTA in several sentence pair matching tasks based on siamese architecture(XX). In order to exclude the influence of the  vector numbers on the recall quality, our base model also groups the click of each document in the same way and may generate multi vectors each document, so that the number of document vectors is the same as the treatment.

\para{MIRA variants.} Besides, to demonstrate the effectiveness of each component in our model, we also take our model's variants  as baseline. One is the model  with the graph neighbours generated in the traditional co-click graph, denoted as MIRA-tradition.  Another two are about attention: one  disables the attention mechanism in graph attention network, denoted as MNIRA-NoAttention, the other takes original GAT as attention mechanism, denoted as MIRA-GAT. This two components are the key to handle co-click noises.

\subsection{Training details}
To fully evaluate our model, we split our dataset into training, validation and test phases.
For Tiangong-ST dataset,  we use 80K clicked query-document pairs as training set. As for  validation set, firstly we  randomly sample 100 queries which have human judged non-bad documents, then for each query, we randomly sample 1000 other documents from all corpus as bad candidates together with their judged non-bad documents as good ones. In validation phase, we choose our model checkpoints and hyper-parameters by $NCG@5$ recalled documents from the candidates for each query. In last test phase, another 500 new queries are used  to evaluated the recall quality from candidates which are generated the same way as in validation phase.

As for Bing dataset, to better demonstrate our model's performance in industrial scenario, we enlarge the data size in three phases: 40 million query-document pairs in training and 1000 queries with 7000 candidates each in validation. Specifically in the test,  we use another 3000 queries with a brute force method where we calculate the vector similarities between each query and  2.2 billion candidate(sampled from total 42B documents) documents to generate final retrieval documents.

We use three layer transformers as our Bert encoder and initialize it with the first three layer of Bert-Base model. As for graph side, we sample 2 neighbours for each node as local neighbours and totally sampling two layer neighbours for one document as calculation subgraph. Then we use a two-layer graph convolutional network containing 768 hidden units to extract neighbour features. The model is trained using the Adam~\cite{kingma2014adam} optimizer with learning rate $8e-5$ and the mini-batch size is set to be 512. In training process, the click query-document pairs are used as positive data with in-batch negative sampling ~\cite{gillick2019learning}. Specifically for each query, we select another document in this mini-batch which has a relative high similarity to this query as a negative sample with a multinomial sampling . By doing so we can not only reuse the document embedding in each mini-batch thus saving training cost, but also enable our model to learn better by not providing too easy cases.

\subsection{Experiment Result}
\begin{table*}[t]\caption{two types of win cases in Bing dataset}
\centering
\setlength{\tabcolsep}{1mm}{
\begin{tabular}{|c|c|c|l|c|}
  \hline
  %&\multicolumn{4}{|c|}{Bing Dataset}& \multicolumn{2}{|c|}{Tiangong-ST Dataset}\\ \hline
  %&\multicolumn{2}{|c|}{Head} &\multicolumn{2}{|c|}{Tail}&\multicolumn{2}{|c|}{ } \\ \hline
   Case Type& Query &\multicolumn{2}{|c|}{Document text} & $\bar{\alpha}$ \\ \hline
   \multirow{3}{*}{\tabincell{c}{key concept \\ enhancing}} & \multirow{3}{*}{fm database}& main text & \tabincell{l}{football manager 2018 player database created by fm inside net  are\\ you looking for the best wonderkids in football manager 2018 \\ football  manager football  manager 2018  3000  football manager 2018 \\ football manager databases} &  \\
   \cline{3-5} & & neighbour 1 text & \tabincell{l}{2000 2001 \colorbox{blue!30}{database}	http www \colorbox{blue!30}{fm} base co uk forum football  manager \\ 2010 editing  32579 2000 2001 \colorbox{blue!30}{database} html	2000 2001 final \colorbox{blue!30}{database}  \\ football manager  2000 2001 \colorbox{blue!30}{database} for \colorbox{blue!30}{fm} 18 football  manager \colorbox{blue!30}{database} \\ 2016 season 2000 2001 football manager \colorbox{blue!30}{databases}  2001  football manager} & 0.44 \\
   \cline{3-5} & & neighbour 2 text & \tabincell{l}{football manager \colorbox{blue!30}{database} 2001 websites \colorbox{blue!30}{fm} base co uk \colorbox{blue!30}{fm} 2018  blog \\football manager 2018 https keywordspace com football  manager \\ \colorbox{blue!30}{database} 2001	football manager \colorbox{blue!30}{databases} 2001} & 0.56 \\   \hline

   \multirow{3}{*}{\tabincell{c}{query term \\ supplement}} & \multirow{3}{*}{\tabincell{c}{how would \\ lines of credit \\ help business}}& main text & \tabincell{l}{https www kabbage com greenhouse article everything you need to  \\know about small business funding	everything you need  to know about \\small business funding	small business funding } &0 \\
   \cline{3-5} & & neighbour 1 text & \tabincell{l}{https www allbusiness com whats a business \colorbox{red!30}{line of credit} and \\ how does it work 15347107 1 html what s a business  \colorbox{red!30}{line of credit} \\ and how does it work allbusiness com what is the  maturity  date \\ for a \colorbox{red!30}{line of credit}	small business funding} & 0.81 \\
   \cline{3-5} & & neighbour 2 text & \tabincell{l}{http bizcap com services asset based lending accounts receivable \\ financing accounts receivable financing accounts receivable financing \\ acquisition small  business funding} & 0.19 \\   \hline

  %Key concept enhance & fm database	&football manager 2018 player database created by fminside net https fminside net football manager 2018 player database	new football manager football database football manager 2018 3000 data base transfer databases football manager 2018 football manager database & 2000 2001 database	http www fm base co uk forum football manager 2010 editing 32579 2000 2001 database html	2000 2001 final database fotboll manager 2000 2001 database for fm18 football manager database  2016 season 2000 2001 football manager databases 2001 football manager 2018 3000 data base & $-21.70\%$ & football manager database 2001 websites fm base co uk fm 2018 blog football manager 2018 https keywordspace com football football manager database 2001	football manager databases 2001 & $52.63$ \\\hline
  %Term complementary & $0.00\%$	&$0.00\%$ & $0.00\%$ & $0.00\%$ & $37.41$ & $62.44$ \\\hline
  %MIG-traditional & $+0.94\%$	&$+1.49\%$ & $+2.73\%$ & $+1.84\%$ & $43.68$ & $83.69$  \\\hline
  %MIG-NoAttention& $+0.41\%$	&$+0.42\%$ & $+1.22\%$ & $+1.18\%$ & $60.31$ & $88.89$ \\\hline
  %MIG-GAT& $+3.02\%$	&$+2.31\%$ & $+3.19\%$ & $+3.28\%$ & $60.43$ & $89.19$ \\\hline
  %MIG &$+4.11\%$ &	$+3.24\%$	&	$+5.10\%$ & $+4.66\%$ & $61.50$ & $92.62$ \\\hline
\end{tabular}}
\label{tab:win_case}
\end{table*}

\para{Retrieval quality analysis.}
For Bing dataset, we split  queries into head and tail based on their frequencies to see how  models perform on hot and long-tail queries and we only report relative change percentage considering the commercial confidentiality. As for Tiangong-ST dataset, we don't distinguish query types due to the small data size.   Table ~\ref{tab:off_result} shows the $NCG$ test results in both datasets.

Firstly  a little surprising , the results of  VPCG is not so competitive, even bad than BM25 in both datasets. We argue that the reasons mainly lie in two aspects: i)for the two query sets we measure on, the click graph degree is a long-tail distribution. For Sogou query set, only 45.7\% queries have more than one clicked documents, and one document just has on average 1.35 associated queries. As for the Bing dataset, 88\% of queries just have only one clicked document, and the average query number  for each document is 4.78. So it means when propagated from document side to query side, we will just get a copy of document vector as query vector at most of the time. The propagation from query side to document side is also affected by this and thus deficient. ii)
In our data set, the proportion of documents and queries that  don't appear in click-graph is significantly larger than the author's data set in the original paper.  For Tiangong-ST and Bing query set, only 83.5\% and 30.4\% of the test queries can be found in the click graph, and only 19.0\% and  19.1\% of  test documents can be found in the click graph, where in VPCG paper these two ratios are 92.5\% and 78.9\% respectively.  The low query and document graph coverage is a disaster for VPCG because those queries and documents couldn't collect and propagate information in the vector propagation phase, instead their vector are generated by the weighted combination of unit vectors, which are not effective. As a matter of fact, in real industrial area a more realistic situation is that a large proportion of documents and queries have not been clicked/appeared in history. So it is important for model  having ability to handle those not in graph queries/documents when leveraging click graph for document retrieval.

Next, we can see that compared with the BM25 model or graph-based methods, the Bert-based deep model improves a lot due to its big step from traditional term-based retrieval to powerful semantic match retireval. Furthermore, on top  of Siamese Bert model, our proposed method can further achieve better retrieval quality. In Bing dataset, for head query our model can achieve $4.11\%$ and $3.24\%$ improvement at NCG@20 and NCG@80 respectively. And for tail queries, the increases expand to $5.1\%$ and $4.55\%$.When it comes to Tiangong-ST dataset, the MIRA  achieves 19.09 gains at NCG@5 and 24.18 gains at NCG@10 compared with the Bert base model.
These results indicate the advantage of MIRA using co-click information to improve document retrieval. Different from graph based methods like VPCG, our model not only leverages  GAT to effectively extract features from neighbours, but also handles the documents  without neighbours well with the help of Bert.
Besides, we notice that the improvement in Tiangong-ST is larger than that in Bing dataset due to the data scale: firstly for Tiangong-ST dataset, its training data size is much smaller than Bing, and thus the model's training is not sufficient to fully understand the queries and documents, so in this case the co-click information is more useful. More importantly, as a web-scale dataset, the document candidate number for each query in Bing is million times than that in Tiangong-ST, as a result,it is more difficult to improve the recall metrics. And the information brought by co-click is more likely to introduce noises in such a large data scale.
Another interesting finding is that in Bing dataset, the tail queries get more benefit from MIRA.  It is because long-tail queries are often more difficult for model to understand and find match terms in document stream, so the information introduced by co-click relations could be more helpful.

\para{Ablation study}
Next, to demonstrate the effectiveness of each component in our model, we compare our framework with its variants and show the results in Table ~\ref{tab:off_result}. Generally speaking, once  disabling one experimental design, the model variant will lose on both data sets. Specifically,in Bing dataset, without multi intention co-click graph, the MIRA-traditional  reduces the gains which MIRA can bring by $60\%$ in NCG metrics. This reveals the  effectiveness of our proposed MICG in handling multi-intention noises. Besides, the attention-mechanism plays an important role in our framework, because it can help model to  better aggregate neighbour information with different weights  thus eliminating the \textit{false click} and \textit{semantic difference} noises.  We can see the MIRA-NoAttention only has a little improvement compared with siamese-Bert model, while  MIRA-GAT can achieve more gains. The last but not the least, the comparison of  MIRA and MIRA-GAT demonstrating that  involving two-factor attention based on GAT  to emphasize the role of exact term matches in semantic query-document matching can indeed  bring  decent NCG gains ($45\%$ improvements).   Another observation is the benefits of these designs are more evident in  industrial-scale dataset and the reason is similar as that in retrieval quality analysis, where the large dataset contains more noise in the co-click relationship, so there is a stronger demand for these designs.

\subsection{Case Study}

In this section, we show some MIRA win cases compared with the siamese base model in  Bing dataset to explain how co-click relation information could improve document retrieval quality.
For simplicity, we only demonstrate the first layer neighbours for each case with their corresponding importance coefficients. Here the  importance coefficient between each node and its neighbour is calculated by the average of their multi-head coefficients $ \bar{\alpha} = \frac{\sum_{m=1}^{M} \alpha^{m}_{ij}}{M}$. And each document text is the concatenate of its multi-source stream(e.g, url, title, anchor and click).
The cases are shown in Table ~\ref{tab:win_case}, and we can see that the MIRA can achieve improvements in two ways, key concept enhancing and  query term complementary.

\para{key concept enhancing.} Neighbor information may contain the key concepts in the main text, and thus it can help model better capture the key information of the original document, excluding the noise information introduced by  other texts. As a result, the final vector can better represent  the meaning of the document.  As shown in the table, the query is "fm database", which contains two concepts "fm" and "database". However, the good document  contains a lot of other concepts , e.g. "fm inside net" and "best wonderkids" . This information has nothing todo with the core content of the document itself and  will add bias to the document's vector  in the vector space if  model couldn't distinguish them from the key concepts. When the neighbor information helps, the situation becomes very different. Both the two neighbor texts we introduced through the co-click relationship contains a lot of term text "fm" and "database"(blue color), especially the "database" that appears only twice in the original text. Thus their term frequencies are improved, so the model can identify them easily and  the final vector is more in line with the true meaning of the document. It is also worth mentioning that these two neighbor documents contribute almost the same to the main document, so the weight assigned to them by the attention mechanism is not much different.

\para{query term supplement.} Neighbor information may also provide some terms or phrases that don't appear in original document text but are  exact matches  with the query.
As we have discussed before, it is more important to focus on exact matching in document retrieval. The original document text and the neighbor text are highly correlated, but they may have different expressions. Therefore, the terms or phrases in the neighbor text can be used as a supplement to the original text when matching with the query. As the case demonstrated, the main document is mainly about "small business funding". It contains the content which is related to the query in the body so it is human-labeled as a good document. But this match information couldn't reflect from its multi-source text. And the semantic correlation between query and text is somehow too weak for model to capture. However, with the help of co-click information, the terms "line of credit" and "business" (red color) in the first neighbour  text serve as a bridge between query and the original good document. Besides, we can see that the two-factor attention also helps by giving more weight to the first neighbour, because compared with the second one which is mainly about "accounts receivable financing", it is more related to the original document.

\section{CONCLUSIONS}
In this work, we study the problem of leveraging co-click information in deep recall model for industrial web search. We propose the Multi-Intention Co-click Graph(MICG), which builds the co-click edges between documents on the  click-intention level based on the history search log. It can provide document neighborhood connections without multi-intention noise and is scalable for billions of nodes. Then we further present an encoding framework MIRA, which leverages Bert and GAT to encode the query and document separately into latent space for semantic matching. With the well designed components, our framework can effectively extract neighbour features for  query and document matching and handle the documents which don't have co-click neighbours well.
We test the proposed framework on one public real search engine dataset Tiangong-ST from Sougou.com and one web-scale real search dataset from Bing.
Our extensive experimental analysis shows MIRA significantly
outperforms several baselines  including term based retrieval model, graph based method and siamese Bert model. A further case study reveals that the improvement from co-click information mainly comes from two aspect, key concept enhancing and query term complementary.

To the best of our knowledge, this is the first work collaborating co-click relations with deep neural networks  to help improving document retrieval quality in industrial-scale web search. Our work uncovers not only the potential, but also the  feasibility to leverage additional information other than the document itself to help encoding  and result in better retrieval metrics.

\bibliographystyle{ACM-Reference-Format}
\bibliography{acmart}

\end{document}